\documentclass[preprint,12pt]{elsarticle}

\usepackage{amssymb}
\usepackage{amsfonts}
\usepackage{amsbsy}
\usepackage{amsmath}
\usepackage{xcolor}
\usepackage{soul}
\usepackage{comment}

\DeclareMathOperator{\diag}{diag}
 \newtheorem{definition}{Definition}[section]
\newtheorem{thm}{Theorem}[section]
\newtheorem{lem}[thm]{Lemma}
\newtheorem{example}[thm]{Example}
\newtheorem{rem}[thm]{Remark}
\newcommand{\pf}{\paragraph{Proof}}
\newcommand{\pfend}{\par\vspace{2ex}\noindent}

\newcommand{\C}{{\cal C}}

\newcommand{\F}{\mathbb{F}}

\newcommand{\beq}{\begin{equation}}
\newcommand{\eeq}{\end{equation}}
\newcommand{\bmat}{\left( \begin{array}}
\newcommand{\emat}{\end{array} \right)}

\newcommand{\AND}{\;\mbox{and }}
\newcommand{\FOR}{\;\;\mbox{for }}

\newcommand{\IM}{\;\mbox{im }}

\begin{document}

\begin{frontmatter}



\title{Easy repair via codes with simplex locality} %% Article title


\author{M.\ Kuijper, J. Lieb and D. Napp}

\address{M.\ Kuijper is with the Department of Electrical and Electronic Engineering, University of Melbourne, VIC 3010, Australia, mkuijper@unimelb.edu.au \\
J.\ Lieb is with the Department of Mathematics, TU Ilmenau, Germany, julia.lieb@tu-ilmenau.de \\
	D. Napp is with the Deptartment of Mathematics, University of  Alicante, Spain, diego.napp@ua.es \\
}

%\affiliation{organization={},%Department and Organization

%% Abstract
\begin{abstract}
In the context of distributed storage systems, locally repairable codes have become important. In this paper we focus on codes that allow for multi-erasure pattern decoding with low computational effort. Different optimality requirements, measured by the code's rate, minimum distance, locality, availability as well as field size, influence each other and can not all be maximized at the same time. We focus on the notion of easy repair, more specifically on the construction of codes that can repair correctable erasure patterns with minimal computational effort. In particular, we introduce the easy repair property and then present codes of different rates that possess this property. The presented codes are all in some way related to simplex codes and comprise block codes as well as unit-memory convolutional codes. We also formulate conditions under which the easy repairs can be performed in parallel, thus improving access speed of the distributed storage system.
\end{abstract}


% \begin{highlights}
% \end{highlights}

%% Keywords

\end{frontmatter}

\section{Introduction}\label{sec_intro}

In today's data-driven world, managing and securing vast amounts of information is crucial. A distributed storage system is a system where data is distributed  across multiple storage units (nodes) which are interconnected over a network,
as we witness in some peer-to-peer (P2P) storage systems \cite{Wuala} and data centers \cite{Azure} that comprise the backbone infrastructure of cloud computing. We call such systems Networked Distributed Storage Systems (NDSS). This approach leverages a network of linked computers or nodes to facilitate data management, access, and distribution. By enabling simultaneous retrieval from different nodes, it improves data availability, redundancy, and access speed.

A fundamental issue that arises in this context is the so-called \emph{Repair Problem}: how to maintain the encoded data when failures (node erasures) occur. When a storage node fails, information that was stored in the node is no longer accessible. As a remedy a node is then added to the
system to replace the failed node. The added node downloads data from a set of appropriate and accessible nodes to recover the information stored in the failed node. This is called {\em node repair}. To
assess the performance of this repair process, there are several metrics that can be considered:
{\em storage cost}, measured as the amount of data stored in the node, {\em repair bandwidth}, measured as the total
number of bits communicated in the network for each repair and {\em locality}, measured as the number of
nodes needed for each repair. For instance, $(n,k)$ maximum distance separable (MDS) codes are
optimal in terms of storage cost since any $k$ nodes contain the minimum amount
of information required to recover the original data. However, to repair one single node it is
necessary to retrieve information from all $k$ nodes. More specifically, repair is achieved by re-encoding the information from these $k$ nodes and storing part of the re-encoded data in the new node.
This results in a poor performance with respect to repair bandwidth as well as locality.
The repair bandwidth metric and the storage cost metric are well-understood metrics, see for example~\cite{Dimakis1,TamoBarg}. Using network coding techniques, several code constructions have been presented that show optimality with respect to repair bandwidth and storage cost, see \cite{Dimakis2} and
references therein. Locality is an important metric that was studied independently by several authors, see \cite{Oggier1}, \cite{Gopalan} and \cite{Dimakis3} among many others,
and it is considered to be one of the main repair performance bottlenecks in many NDSS, e.g., in
cloud storage applications.

When a code of locality $r$ is used, then one needs to contact at most $r$ nodes to repair one node. In this paper this will be referred to as $r$-easy repair. The case of {\em multiple} node erasures is more involved as  a single subset of nodes can repair a particular piece of redundant data and therefore if a node from this repair subset is also not available, data cannot be repaired “locally”, increasing the cost of the repair. Hence, it is desirable to obtain codes providing multiple repair alternatives per node. Some interesting  results on this problem were first presented in \cite{PraKam}, seeking to extend the ideas in \cite{Gopalan} for one erasure. General and optimal constructions of locally repairable (convolutional) codes over finite fields $\F_{q^m}$ can be found in \cite{Chen,umbertonapp,Dimakis3,TamoBarg} and over small finite fields in \cite{Blaum2,Huang,binaryLi,Natalia}. It is also worth mentioning the results in \cite{m,Jingxue, calderbank,colweight,Oggier1,Oggier2} where some code schemes, akin to the first one presented here, are introduced. A preliminary version of this work was presented in \cite{MargretaDiego}.

In this paper, we study two different coding scenarios: when the data is stored in one single vector of fixed length and when the data is stored in a long sequence of data vectors. The first scenario is the one typically considered in the literature, leading to the use of block codes. However, with streaming applications in mind, one could argue that a more efficient approach to repairing erasures is to rely either on the symbols at a specific time instant or on a small group of symbols ''close" to the erasure within the stream. This leads us to investigate also the use of convolutional codes for storage. Unlike block codes, where grouping and length are fixed, the flexibility obtained by using sliding window convolutional decoding allows us to slide along the transmitted sequence and adapt the decoding to the erasure pattern.

We introduce block codes and unit-memory convolutional codes based on \emph{simplex codes} and \textit{Low Column Density Generator Matrix} (LCDGM). The codes are defined over $\F_{2^m}$ 
and share some locality properties that allow us to repair all the erased nodes by performing few XOR operations, even in the presence of multiple erasures. As these operations are computationally very efficient, we call them \textit{easy repairs}.

In this work we focus on codes especially suitable when locality is relevant, multiple erasures may occur and repairs can be performed by easy repairs or even \textit{parallel} easy repairs. This is related to locally repairable codes with $(r,t)$-availability, see for instance \cite{Guruswami,Malluhi,Kumar2015,Natalia,Rawat,TamoBarg}. However, these works do not focus on easy repair and their optimal codes require relatively large finite fields, e.g. they are based on Reed-Solomon \cite{TamoBarg} or Gabidulin codes \cite{Rawat}. In this paper we emphasize the easy repair angle as well as the parallel repair angle as these are
highly relevant for computational feasibility as well as access speed.
We also note that in this context convolutional codes have been much less investigated, see for instance \cite{Chen,Zyablov,umbertonapp,umbertonapp2,Zhu}. In this paper we  focus on convolutional codes tailored to easy repair as well as parallel repair of erased nodes.

After introducing the notion of locality of codes and defining several types of easy repairs, we present the locality properties of the binary simplex code in Section \ref{sec_block}. We recall from our preliminary work~\cite{MargretaDiego} that if no more than $\frac{n-1}{2}$ erasures occur, then we can easily recover all nodes by XORing two live nodes at the same time in a parallel recovery fashion. Moreover, if an erasure pattern is correctable, then we can recover all nodes by easy repairs in a sequential recovery fashion. We call this last property the \emph{easy repair property}.

The drawback of using simplex codes as described in Section 3 are the limited values of the code rate we can cover with these codes. Hence, in the following sections we  present constructions with different rates that still retain the \emph{easy repair property}. To this end we investigate punctured simplex block codes in Section 4.
Finally, in Section \ref{sec_conv} we present a class of convolutional codes that can encode long sequences of data vectors and has some interesting locality properties.
In particular, these codes also have the easy repair property.
Moreover, depending on the number of erasures, the proposed convolutional code can repair each node
in parallel performing just XORing of between 2 and 5 nodes.
In Section 6, we compare the performance of all the constructions presented in the paper. We close the paper with conclusions in Section \ref{sec_conclusions}.

\section{Locality and Easy Repairs}

Let $q=2^m$ and let $\F_q$ be a finite field with $q$ elements. A $q$-ary linear $(n,k)$ block code $C$ of length $n$ and rank $k$ is a $k$-dimensional linear subspace of $\F_q^n$. Full-rank matrices $G\in \mathbb{F}_q^{k \times n}$ and $H \in \mathbb{F}_q^{n \times (n-k)}$ with the property that
\begin{eqnarray*}
{C}  &=& \IM_{\F_q}G = \left\{ c= u G \in \mathbb F_q^{n}:\, u \in \mathbb F_q^{k}\right\} \\
&=& \ker_{\F_q}H = \left\{ c  \in \mathbb F_q^{n}:\, Hc^\top=0 \right\},
\end{eqnarray*}
are called \textit{generator matrix} and \textit{parity-check matrix} of $\C$, respectively.\\

The early literature on distributed storage focuses on the notion of locality, formally defined below, which is relevant to the decoding of so-called single-erasure patterns, i.e. erasure patterns with up to one erasure per word.
%\vspace{.4cm}
\begin{definition}\label{def-locality}
An $(n,k)$ code has {\em locality} $r$ if each codeword symbol in a codeword is a linear combination of at most $r$ other symbols in the codeword.
\end{definition}

It was shown in~\cite{Gopalan} (see also~\cite{TamoBarg}) that there exists a natural trade-off among redundancy, locality and code minimum (Hamming) distance:
\begin{thm}
Let $C$ be an $(n,k)$ linear block code with minimum distance $d$ and locality $r$. Then
$$
n-k +1 - d \geq \left\lfloor \frac{k-1}{r} \right\rfloor .
$$
\end{thm}

From the above bound it is seen that MDS codes do not perform well with respect to locality. Indeed, since the gap $n-k+1-d$ equals zero they have only trivial locality $r=k$.
  When the gap $n-k +1 - d$ is nonzero then it can be "used" for locality purposes. The class of locally repairable codes (LRC)~\cite{m,Guruswami,Huang,KamPra,Oggier1,Dimakis3}, addresses the repair problem focusing on minimizing the number of nodes contacted during the repair process of single-erasure patterns. 

For the case in which multiple erasures occur we have the following definition which generalizes Definition~\ref{def-locality}, see also~\cite{Rawat}.

\begin{definition}\label{def-availability}
An $(n,k)$ code has {\em $(r,t)$-availability} if for each codeword symbol $\hat{c}$ there exist $t$ disjoint groups each consisting of at most $r$ symbols different from $\hat{c}$ and, for each of these $t$ groups, $\hat{c}$ can be written as a linear combination of the symbols in this group.
\end{definition}

 When an $(n,k)$ code $C$ is used for distributed storage, a data object or file $u=(u_1,u_2,\dots,u_k)\in \F_q^k$ of $k$ symbols is redundantly stored across $n$ different nodes in $c=(c_1,c_2,\dots,c_n)= u G\in \F_q^n$, where $G$ is the generator matrix of $C$. Hence, nodes correspond to columns of $G$. 

In this paper we set $q=2$ and propose several binary erasure codes for distributed storage. All of these can be used in a straightforward way to store data objects consisting of symbols from binary extension fields $\F_{2^m}$.

Consider a binary $(n,k)$ code whose nodes are affected by erasures. Let $S= \{c_1,\dots,c_n  \}$ be the set of nodes, $S^e\subset S$ the set of erased nodes
and $S^\ell\subset S$ the set of live nodes (=non-erased). %and let $S_i^e= \{ i \ | \ c_i \in S^e \}$ and $S_i^\ell= \{ i \ | \ c_i \in S^\ell \}$  be the indices of the erased and live nodes, respectively.
A node $c_i$ is said to be \textit{related} to $\gamma$ nodes $c_{j_1}, \dots ,c_{j_{\gamma }}$  if $c_i=c_{j_1} + \cdots + c_{j_{\gamma}}$. If $c_i\in S^e$ is related to $c_{j_1}, \dots ,c_{j_{\gamma}}$ where $c_{j_k} \in S^\ell$ for $k=1,\dots , \gamma$, then $c_i$ can be recovered by performing $\gamma-1$ XORing operations.
\begin{definition}
We say that $c_i\in S^e$ allows for \emph{$r$-repair} if there exists $\gamma \leq r$ such that $c_i=c_{j_1} + \cdots + c_{j_{\gamma}}$ with $c_{j_k} \in S^\ell$ for $k=1,\dots ,\gamma$.
\end{definition}
\begin{definition}
If an erasure pattern can be decoded by only using $r$-repairs then we say that the erasure pattern allows for $r$-repair. When each of the erased nodes in an erasure pattern allows for $r$-repair then we say that the erasure pattern allows for \emph{parallel $r$-repair}.
\end{definition}

\begin{thm}\label{thm_availability}
Let $C$ be an $(n,k)$ code with $(r,t)$-availability. Then any erasure pattern with $\leq t$ erasures allows for parallel $r$-repair.
\end{thm}
\pf
Consider an erased node $c$. Since $|S^e| \leq t$ there are only $\leq t-1$ other erased nodes. So at least one of the $t$ groups related to $c$ must consist entirely of live nodes. As the size of this group is upper bounded by $r$, it follows that $c$ can be reconstructed by $r$-repair. This concludes the proof.
\pfend
We note that erasure patterns that do not allow for parallel $r$-repair may still allow for $r$-repair, namely in a sequential way. 

In this paper we will mostly focus on the case $r=2$, i.e., decoding via either replication or a simple XOR of two live nodes.

\begin{definition}\label{defEasy}
We refer to $r$-repair for $r=2$ as \emph{easy repair}. We say that a code has the \emph{Easy Repair Property} if any correctable erasure pattern allows for easy repair.
\end{definition}
In this paper we consider two types of codes for distributed storage: block codes and convolutional codes. We mainly focus on their easy repair capabilities. More specifically, we first consider the simplex code and then introduce several other binary block codes. For the more involved convolutional case we propose a unit memory convolutional code based on simplex codes and demonstrate that it has the easy repair property. Furthermore, we investigate its parallel $r$-repair properties for $r=3,4$ and $5$.

\section{Simplex block codes for distributed storage }\label{sec_block}
In this section we recall several results from our earlier 2014 work~\cite{MargretaDiego} on the easy repair properties of binary simplex codes and rephrase them in terms of more recent concepts such as $(r,t)$-availability, thus providing a foundation for the later sections of this paper.

\begin{lem}\label{le:aux}
Let $C$ be a linear $(n,k)$ code and let $S^e$ be the set of indices of erased nodes. 
Let $|S^e| = n-s$. Then the following statements are equivalent:
\begin{enumerate}
  \item The erasure pattern corresponding to $S^e$ is correctable;
  \item The $k \times s$ matrix $\hat G$ formed by deleting the $i$-th columns of $G$ where $i\in S^e$, is right invertible;
  \item The $(n-k) \times (n-s)$ matrix $\hat H$ formed by the $i$-th columns of $H$ where $i\in S^e$, is left invertible.
\end{enumerate}
\end{lem}

\pf
Clearly 1) holds if and only if there do not exist two different codewords that coincide in positions corresponding to $S^\ell$, the set of indices of live nodes. Since the code is linear, this is equivalent to the non-existence of a nonzero codeword whose symbols at positions in $S^\ell$ are zero. The latter is clearly equivalent to the linear independence of the columns of $\hat H$, thus proving the equivalence of 1) and 3). Next, we prove the equivalence of 1) and 2). Write $S^\ell = \{ j_1 , \ldots , j_s\}$. Consider the system of equations
 \[
 \bmat{ccc} c_{j_1} & \cdots & c_{j_s}\emat= u \hat G .
   \]
Solving this system is equivalent to repairing all erasures, i.e. the recovery of $u$. The equivalence of 1) and 2) now follows from the fact that this system is solvable if and only if $\hat G$ is right invertible.
This concludes the proof.
\pfend

Next we recall the definition of a binary simplex block code and study its properties when used for distributed storage.

\begin{definition}
Let $k$ be a positive integer, $n= 2^k -1$ and let $G$ be a $k  \times  n $ matrix whose columns are the distinct non-zero vectors of $\F^k_2$. Let $C$ be the binary block code over $\F_2$ that has $G$ as its generator matrix. Then $C$ is called a binary \textit{simplex} $(n,k)$ block code.
\end{definition}

Binary simplex block codes are classical codes that are dual to the binary Hamming codes. Their minimum distance equals $d = 2^{k-1} $ which implies that their erasure correcting capability equals
\[
d-1 = 2^{k-1}  -1 = \frac{n-1}{2},
\]
i.e.~any erasure pattern with $\leq \frac{n-1}{2} $ number of erasures is correctable.

\begin{example}\label{ex:01}
Let $k=3$ and $C\subset \F_2^7$ be a simplex $(7,3)$-code. Then, its generator matrix $G$ and parity-check matrix $H$ are given by;
$$
G= \left(
     \begin{array}{ccccccc}
       1 & 0 & 0 & 1 & 1 & 0 & 1 \\
       0 & 1 & 0 & 1 & 0 & 1 & 1 \\
       0 & 0 & 1 & 0 & 1 & 1 & 1 \\
     \end{array}
   \right)
$$
and
$$ \ H = \left(
                    \begin{array}{ccccccc}
                       1 & 1 & 0&1 & 0 & 0 & 0\\
                       1 & 0 & 1&0 & 1 & 0 & 0\\
                       0 & 1 & 1&0 & 0 & 1 & 0\\
                       1 & 1 & 1&0 & 0 & 0 & 1\\
                    \end{array}
                  \right),
$$
respectively.
\end{example}

 The binary simplex code possesses several good locality properties, starting with the next lemma.

\begin{lem}\label{lem:01}
The $(n,k)$ binary simplex code has $(2,\frac{n-1}{2})$-availability.
\end{lem}
\pf
Denote its set of nodes by $S= \{c_1,\dots,c_n  \}$. Choose any node, say $\hat c\in S$. We need to show that $\hat c$ is related to $\frac{n-1}{2}$ different pairs. Let $\hat S :=\{ \hat c\}$. Take $c_{1_j}\in S \setminus\hat S$ and set $c_{1_k} = c_{1_j} + \hat c$. Due to the fact that any sum of two columns of $G$ is another column of $G$ we have that for any $c_j,c_k\in S$, $c_j+c_k \in S$. Hence, $c_{1_k}\in S$ and let $\hat S := \hat S \bigcup \{ c_{1_j}, c_{1_k}\}$. Again take any node $c_{2_j}\in S \setminus\hat S$ and set $c_{2_k}= c_{2_j} + \hat c \in S$. Note that $c_{2_k}\notin \hat S$ and therefore the pairs $(c_{1_j},c_{1_k})$ and $(c_{2_j},c_{2_k})$ are disjoint. Let $\hat S := \hat S \bigcup \{ c_{2_j}, c_{2_k}\}$ and the cardinality of the set $\hat S$ is increased by two in each step. Repeating this process $\frac{n-1}{2}$ times, we obtain $\frac{n-1}{2}$ disjoint pairs related to $\hat c$. Since the choice of $\hat c$ is arbitrary, this concludes the proof.
\pfend
\begin{thm}\label{th:01}
Let $C$ be an $(n,k)$ binary simplex code. Then any erasure pattern whose number of erasures is within the erasure correcting capability of $C$ allows for parallel easy repair.
\end{thm}
\pf
The statement follows immediately from the previous lemma and Theorem~\ref{thm_availability}.
\pfend

We recall that for the $(n,k)$ simplex code the erasure correcting capability is $\frac{n-1}{2}$. We now turn our attention to the larger class of erasure patterns that are correctable and possibly have $> \frac{n-1}{2}$ erasures.

\begin{lem}\label{exists}
Let $C$ be an $(n,k)$ binary simplex code. Then for any correctable erasure pattern, there exists an erased node that allows for easy repair.
\end{lem}

\pf
Denote the set of erased nodes by $S^e$ and the set of live nodes by $S^\ell$; denote the set of its indices by $S_i^\ell=\{ j_1 , \ldots , j_s\}$.
Consider the system of equations
 \[
 \bmat{ccc} c_{j_1} & \cdots & c_{j_s}\emat= u \hat G ,
  \]
where $\hat G$ is the $k \times s$ matrix that remains after deleting from $G$ the $i$-th columns at erased positions. Since $S^e$ corresponds to a correctable erasure pattern, this system of equations is solvable over $\F_2$. Thus it follows from Lemma~\ref{le:aux} that $\hat G$ has rank $k$. As a result, for all $\hat c\in S^e$ there exist an integer $g$ and $a_j\in S_i^\ell$ for $j=1, \dots, g$ such that
$$
\hat c= c_{a_1} +\dots + c_{a_g}.
$$
Consider the vector $c_{a_1} + c_{a_2}$ and denote this vector by $c_{b_1} $.
Then $c_{b_1}$ must be a column vector of $G$ because $G$'s columns comprise all nonzero vectors in $\F^k_2$. If $c_{b_1} \in S^e$, then we have found an erased node that is easily repairable. If not, then as a column vector of $G$, $c_{b_1}$ must be a live node $ \in S^\ell$. We have $\hat c= c_{b_1} + c_{a_3} +\dots + c_{a_g}$. Again, if $c_{b_1} + c_{a_3} \in S^e$, then we have found one erased node that is easily repairable. If not, {\em i.e.}, if $c_{b_1} + c_{a_3} \in S^\ell$, then denote $c_{b_2}=c_{b_1} + c_{a_3}$ and therefore $\hat c= c_{b_2} + c_{a_4} +\dots + c_{a_g}$. This process must end yielding either that $c_{b_j} + c_{a_{j+2}} \in S^e$ with $c_{b_j}, c_{a_{j+2}} \in S^\ell$ for some $j\in\{ 1,2, \dots, g-3\}$ or $\hat c= c_{b_{g-2}} + c_{a_g}$. In both cases we obtain an easy repair and the proof is completed.
\pfend

\begin{thm}\label{th:02}
An $(n,k)$ binary simplex code has the Easy Repair Property, as defined in Definition~\ref{defEasy}.
\end{thm}
\pf
Because of Lemma \ref{exists}, there exists an erased node that allows for easy repair. Once repaired, the erasure pattern is obviously still correctable. As a result, we can apply Lemma \ref{exists} again leading to yet another erased node that allows for easy repair. Continuing these easy repairs over and over we proceed until all erasures are recovered.
\pfend

%\vspace{.2cm}

\begin{example}
Consider the matrices $G$ and $H$ as defined in Example \ref{ex:01}. Suppose that we have a file $u\in \F_2^3$ to be stored in $7$ nodes, \emph{i.e.}, $u G= c=(c_1,\dots,c_7)\in \F_2^7$, and that erasures occur in nodes $c_1,c_2,c_4$ and $c_6$, \emph{i.e.}, $S^e= \{ 1,2,4,6\}$, $S^\ell= \{ 3,5,7\}$.
Thus the pattern is correctable despite the fact that the number of erasures is outside of the erasure correcting capability. It now follows from Lemma~\ref{exists} that there exists an erased node that allows for easy repair. Indeed, $c_1=c_3 + c_5$ and in fact there exist several erased nodes that allow for easy repair, namely also $c_2= c_5 + c_7$ and $c_4= c_3 + c_7$. By Theorem~\ref{th:02}, we can repair all nodes by easy repairs. Indeed, $c_2$ and $c_4$ already allow for easy repair and, once $c_2$ is repaired, we repair $c_6$ from $c_6=c_2+ c_3$.\\
Note that the node $c_6$ cannot be the first node to be easily repaired and we need to repair a different node first. However, when the number of erasures does not exceed the erasure correcting capability of the code, then by Theorem~\ref{th:01} parallel easy repair is possible. Thus \emph{any} erased node can be chosen to start the repair.
\end{example}

\section{Low Density Generator Matrix block codes for distributed storage }\label{sec_pun}

In this section we first investigate puncturing of the simplex codes of the previous section in order to derive a higher rate code that still allows for easy repair. In particular, we investigate which columns of the simplex generator matrix can be deleted while still preserving easy repair properties. It turns out that the resulting code has a sparse generator matrix and thus belongs to the family of Low Density Generator Matrix (LDGM) codes. These codes offer significant computational advantages, not only for decoding but also for encoding. This is a highly desirable feature when dealing with large datasets, see for instance \cite{Malluhi}.

\begin{definition}\label{C1}

Let $k\in\mathbb N$, $k\geq 2$. Define the $(\frac{k(k+1)}{2},k)$ code $\C_1$ by its generator matrix $[I_k\ \tilde{G}_k]$, where $\tilde{G}_k$ is the $k \times \frac{k(k+1)}{2}$ binary matrix whose columns are all vectors from $\mathbb F_2^k$ that have weight $2$.
\end{definition}

\begin{lem}\label{weight2}
    Each nonzero vector in the rowspan of $\tilde{G}_k$ has weight at least $k-1$.
\end{lem}

\begin{pf}

For $k=2$, we have $\tilde{G}_2=\begin{pmatrix}
 1 \\ 1
\end{pmatrix}$ and the lemma clearly holds for this case.
For $k \geq 3$ one can write $\tilde{G}_k$ recursively as
\begin{eqnarray}\label{recursion}
\tilde{G}_k=\begin{pmatrix}
    1 & \hdots & 1 & 0 & \hdots & 0\\
    & I_{k-1} & & & \tilde{G}_{k-1}
\end{pmatrix}.
\end{eqnarray}
We observe that the first row of the matrix on the right hand side of~(\ref{recursion}) is the sum of its other rows (and this is the only nontrivial linear combination of rows of $\tilde{G}_k$ that equals zero). Therefore a nonzero linear combination of the rows of $\tilde{G}_{k}$ is a linear combination $v$ of the rows of $[I_{k-1}\ \tilde{G}_{k-1}]$. If $v$ has all of its last $\frac{(k-1)(k-2)}{2}$ components equal to zero then it must be the sum of all rows of $[I_{k-1}\ \tilde{G}_{k-1}]$ and thus have weight $k-1$. This is because there is only one nontrivial linear combination of rows of $\tilde{G}_{k-1}$ that equals zero. In the alternative case where not all of the last $\frac{(k-1)(k-2)}{2}$ components of $v$ are equal to zero, it follows per induction that $v$ has weight $\geq k-2+1=k-1$.

\end{pf}

\begin{thm}
    The code $\C_1$ as defined in Definition~\ref{C1} has rate $\frac{2}{k+1}$ and
    minimum distance $k$.
\end{thm}

\begin{pf}
The rate of $\C_1$ equals $\frac{2k}{k(k+1)} = \frac{2}{k+1}$. To show that $d=k$, we first observe that $\C_1$ equals the rowspan of $[I_k\ \tilde{G}_k]$ which equals the rowspan of the matrix $\tilde{G}_{k+1}$. Using the linearity of the code it now follows from Lemma \ref{weight2} that its minimum distance equals $k$.

\end{pf}

\begin{lem}
        The code $\C_1$ as defined in Definition~\ref{C1} has $(2,k-1)$-availability.
\end{lem}

\begin{pf}
    We show that there are exactly $k-1$ ways to write a column of $[I_k\ \tilde{G}_k]$ as the sum of two other columns of this matrix, as follows.
    Take any column $c$ and consider the first position where it has a $1$. There are exactly $k-1$ other columns $c_1$ which have a $1$ at this position. The weight of $c-c_1$ is at most $2$ and hence $c-c_1$ is a column $c_2$ of $[I_k\ \tilde{G}_k]$. Therefore, for each column $c$, there are exactly $k-1$ pairs $(c_1,c_2)$ such that $c=c_1+c_2$. Obviously none of these $k-1$ pairs are intersecting. This concludes the proof.
\end{pf}

The next theorem shows that $\C_1$ inherits a property of the simplex code as it is the analogon of Theorem~\ref{th:01}. Recall that the erasure-correcting capability of $\C_1$ equals $d-1 = k-1$.
\begin{thm}\label{c1-parallel}
Let the code $\C_1$ be defined as in Definition~\ref{C1}. Then any erasure pattern whose number of erasures is within the code's erasure-correcting capability allows for parallel easy repair.
\end{thm}
\pf
The statement follows immediately from the previous lemma and Theorem~\ref{thm_availability}.
\pfend
The next theorem shows that $\C_1$ inherits yet another property of the simplex code, namely Theorem~\ref{th:02}, which is the Easy Repair Property, as defined in Definition~\ref{defEasy}. We note however that the proof of the next theorem is more involved than the proof of Theorem~\ref{th:02}.

\begin{thm}
The code $\C_1$ as defined as in Definition~\ref{C1} has the Easy Repair Property.
\end{thm}
\pf
We first prove that there exists an erased node that allows for easy repair. To see this, choose an arbitrary erased node $\hat c$. Because of the correctability of the erasure pattern it is possible to write $\hat c$ as a sum of live nodes as
$$
\hat c= c_{a_1} +\dots + c_{a_g},
$$
where $g\geq 2$.
If $g=2$ then clearly we have found an erased node that allows for easy repair, namely $\hat c$.
For the case $g\geq 3$ we will now show that there must exist a pair of live nodes in the RHS of the above equation such that their sum is a column of the code's generator matrix, i.e. has weight $\leq 2$. Suppose that this is not true. Then at most one live node in the RHS of the above equation has weight 1.

Also, none of the pairs in the RHS of the above equation have a nonzero coinciding entry. It follows that
\[
wt(\hat c) \geq 2(g-1)+1=2g-1\geq 5.
\]
However, this contradicts the assumption that $\hat c$ has weight $\leq 2$. We conclude that there exists
a pair of live nodes $(c_{a_i}, c_{a_j})$ in the RHS of the above equation such that the sum $c_{a_i} + c_{a_j}$ has weight $\leq 2$, thus is a column in the code's generator matrix. If this sum is an erased node, then we are done. If this sum is a live node, then proceed recursively as in the proof of Lemma~\ref{exists}. We conclude that there exists an erased node that allows for easy repair. Once repaired, the erasure pattern is of course still correctable and we repeat over and over until all erasures are recovered by easy repair. This concludes the proof.
\pfend

We now present a different LDGM block code which we shall call $\C_2$. For $k \geq 4$ the code $\C_2$ has higher rate than the above code $\C_1$ but still possesses some of the easy repair properties as shown below.
Our construction will provide an initial step towards the next section on convolutional codes.

\begin{definition}\label{C2}
Let $k\in\mathbb N$, $k\geq 2$. Let $e_i$ be the $i$-th standard unit vector in $\F_2^k$. Define the $(2k+1,k)$ code $\C_2$ by its generator matrix $G_c$, given by
\begin{eqnarray}
    G_c=\begin{pmatrix}
    e_1 & e_1 & e_1+e_2 & e_2 & e_2+e_3 & e_3 & \cdots & e_{k-1}+e_k & e_k & e_k
\end{pmatrix}.\label{Gc}
\end{eqnarray}
\end{definition}

Clearly, the code $\C_2$ has rate $\frac{k}{2k+1}$ and minimum distance $3$. Unlike $\C_1$ and the simplex code, parallel easy repair for erasure patterns whose number of erasures is within the erasure correcting capability of $\C_2$ is not possible. For parallel repair we have the following lemma and theorem instead.

\begin{lem}
        The code $\C_2$ as defined in Definition~\ref{C2} has $(3,2)$-availability.
\end{lem}

\pf
The columns of $G_c$ are either of the type $e_i$ or of the type $e_i + e_{i+1}$. It is obvious that each column of the type $e_i$ has two disjoint repair pairs, so certainly $(3,2)$-availability. Further, each column of the type $e_i+e_{i+1}$ has two disjoint repair triples because it can be written as $e_i+e_{i+1}=e_{i-1}+(e_{i-1}+e_i)+e_{i+1}$ as well as $e_i+e_{i+1}= e_{i+2}+(e_{i+1}+e_{i+2})+e_i$. This concludes the proof.
    \pfend

Clearly the erasure-correcting capability of $\C_2$ equals $d-1 = 2$.
\begin{thm}\label{c2-parallel}
Let the code $\C_2$ be defined as in Definition~\ref{C2}. Then any erasure pattern whose number of erasures is within the code's erasure-correcting capability allows for parallel $3$-repair.
\end{thm}
\pf
The statement follows immediately from the previous lemma and Theorem~\ref{thm_availability}.
\pfend

The situation for correctable erasure patterns is similar as for $\C_1$ and the simplex code: sequential easy repair is possible as we show next.

\begin{thm}
 The code $\C_2$ as defined as in Definition~\ref{C2} has the Easy Repair Property.
\end{thm}
\pf
We first prove that there exists an erased node that allows for easy repair. To see this, choose an arbitrary erased node $\hat c$. Because of the correctability of the erasure pattern it is possible to write $\hat c$ as a sum of distinct live nodes as
$$
\hat c= c_{a_1} +\dots + c_{a_g},
$$
where $g\geq 2$.
If $g=2$, then clearly we have found an erased node that allows for easy repair, namely $\hat c$.
For the case $g\geq 3$ we will now show that there must exist a pair of live nodes in the RHS of the above equation such that their sum is a column of $G_c$. Suppose that this is not true. Then no column pairs of the form $(e_i , e_i + e_{i+1})$ or $(e_{i+1} , e_i + e_{i+1})$ occur in the RHS of the above equation. It follows that
\[
wt(\hat c) \geq g\geq 3.
\]
However, this contradicts the assumption that $\hat c$ has weight $\leq 2$ as a column of $G_c$. We conclude that there exists
a pair of live nodes $(c_{a_i}, c_{a_j})$ in the RHS of the above equation such that the sum $c_{a_i} + c_{a_j}$ is a column of $G_c$. If this sum is an erased node, then we are done. If this sum is a live node, then proceed recursively as in the proof of Lemma~\ref{exists}. We conclude that there exists an erased node that allows for easy repair. Once repaired, the erasure pattern is of course still correctable and we repeat over and over until all erasures are recovered by easy repair. This concludes the proof.
\pfend

As explained in the next section in Remark~\ref{remarkGc}, the matrix $G_c$ as defined in equation~(\ref{Gc})
can be seen as a
sliding generator matrix of a unit-memory convolutional code with polynomial generator matrix $G(D)=G_0+G_1 D\in\mathbb F_2[D]^{1\times 2}$ where $G_0=(1\ 1)$ and $G_1=(1\ 0)$. 
In general, $G_c\otimes G$ with
a generator matrix $G$ of a $k$-dimensional simplex code can be seen as the sliding generator matrix of a unit-memory convolutional code with polynomial generator matrix $G(D)=G_0+G_1 D\in\mathbb F_2[D]^{k\times (2^{k+1}-2)}$
where $G_0=[G\ G]$ and $G_1=[G\ 0]$.
This idea is properly introduced in the following section, dedicated to (unit-memory) convolutional codes tailored to distributed storage and easy repairs.

\section{Convolutional simplex codes for distributed storage }\label{sec_conv}

In contrast to block codes, binary convolutional codes process a continuous sequence of data vectors $u_{0},u_{1}, \ldots$, $u_i \in \F_2^k$ instead of a single vector. The convolutional encoder encodes each vector $u_i$ at time instant  $i$ in a sequential fashion  to create dependencies between the encoded vectors to cope with erasure patterns. Introducing the notation $D$ for the \emph{delay operator}, we can represent the message sequence $u_{0},u_{1}, \ldots,u_{s}$ as a polynomial  $u(D) = u_{0}+u_{1} D +u_{2}D^2 + \cdots +u_{s}  D^{s} \in \F_2^k[D]$. In this representation the encoding process of convolutional codes, and therefore the notions of convolutional code and convolutional encoder, can be presented as follows \cite{GlRoSm,Encyclopedia}.

\medskip

A binary \textit{convolutional code} ${\mathcal C}$ of rate $k/n$ is an $\F_2[D]$-submodule of $\F_2^n[D]$ of rank $k$ given by a polynomial \emph{encoder matrix} $G(D) \in \F_2^{k \times n}[D]$,
\[
\mathcal{C}  = \IM_{\F_2[D]}G(D) =\{ c(D)=u(D) G(D) : u(D) \in \F_2^k[D] \},
\]
where $u(D) $ is the information vector to be stored. If
\[
G(D) = \sum_{i=0}^\ell G_i D^i \in \F_2^{k \times n}[D]
\]
with $G_\ell\neq 0$, then $\ell$ is called the \emph{memory} of $G(D)$, since it needs to ``remember" the inputs $u_i$ from $\ell$ units in the past. Note that when $\ell=0$, the encoder is constant, so it can be seen as a generator matrix of a block code. Hence, the class of  convolutional codes generalizes the class of linear block codes in a natural way. Convolutional codes with $\ell =1$ are called \emph{Unit Memory (UM) codes}.

For the purposes of this paper it is useful to use matrix notation. Let $c(D)= c_0 + c_1 D + \cdots +c_{s+\ell}D^{s+\ell }$,
\begin{align*}
	c_{\rm total} & := \begin{pmatrix}
		c_0 & c_1 & \cdots & c_{s+\ell}
	\end{pmatrix} \in\F_2^{(s+\ell+1)n}, \\
	u_{\rm total} & := \begin{pmatrix}
		u_0 & u_1 & \cdots & u_{s}
	\end{pmatrix} \in\F_2^{(s+1)k},  \nonumber
\end{align*}
and the \textit{sliding generator matrix} 
\begin{equation}\label{eqG-total}
	G_{\rm total} :=
	\begin{pmatrix}
		G_0 & G_1 & G_2 & \cdots & \cdots & G_\ell & & & & \\
		& G_0 & G_1 & G_2 & \cdots & \cdots & G_\ell &  & &  \\
		& & \ddots & \ddots & \ddots & & & \ddots  \\
		& & &  \ddots & \ddots & \ddots & & & \ddots  \\
		& & & & G_0 & G_1 & G_2 & \cdots &   \cdots & G_\ell
	\end{pmatrix} .
\end{equation}
Note that $G_{total} \in\F_2^{(s+1)k\times(s+\ell+1)n}$ and we have that
\begin{equation}
    c_{\rm total} = u_{\rm total} G_{\rm total} .\label{c-total}
\end{equation}

The \textit{free distance} of a convolutional code $\mathcal{C}$ is defined as $$d_{free}= \min \{wt(c(D))\ |\ c(D) \in \mathcal{C},\ c(D) \neq 0\},$$ where $wt(c(D)) = \sum_{i\in\mathbb N_0} wt (c_i)$ is the Hamming weight of $c(D)$.
While the free distance of a convolutional code is a measure for the total number of erasures such a code can correct, the following notion of column distance captures the correcting capabilities of the convolutional code within intervals (windows). The $j$-th \textit{column distance} of $\cal C$ is defined as
\begin{eqnarray*}
d_j &=& \min\{wt(c_0+ c_1D + \dots + c_j D^j) : \ c(D)=\sum_{i \in \mathbb N_0} c_i D^i\in {\cal C} \mbox{ and } \ c_0 \neq 0  \}.
\end{eqnarray*}

Convolutional codes offer an interesting alternative to block codes especially when there is an undetermined large amount of data that can be encoded sequentially. In this scenario one can recover the erasures by using the sliding window property of convolutional codes, i.e, one can slide along the transmitted sequence and choose where to start the recovery. This is due to the fact that convolutional codes impose redundancy among the blocks along the sequence. 
Thus, the locality properties will depend on the memory of the code, see \cite{Chen,Zyablov,umbertonapp,umbertonapp2,Zhu} for locally repairable convolutional codes with arbitrary large memory. In this section we seek to design unit
memory (UM) codes that allow for sliding window local recovery with easy repairs. \\
We note that for UM codes equation~(\ref{eqG-total}) becomes
\begin{equation}\label{eqG-totalUM}
	G_{\rm total} :=
	\begin{pmatrix}
		G_0 & G_1 &   \\
		& G_0 & G_1 &    \\
		& & \ddots & \ddots  \\
		& & &  \ddots & \ddots   \\
		& & & & G_0 & G_1 &
	\end{pmatrix} ,
\end{equation}
with $G_{total} \in\F_2^{(s+1)k\times(s+2)n}$.

\begin{definition}\label{simplexcc}
 Let $G\in\mathbb \F_2^{k\times \frac{n}{2}}$ be the generator matrix of a simplex code (i.e. $ \frac{n}{2}=2^k-1$). Let $\mathcal{C}$ be the binary $(n,k)$ UM code
  given by the encoder $G(D)=G_0 + G_1 D$ where  $G_0=\begin{bmatrix}
     G & G
 \end{bmatrix}\in\mathbb \F_2^{k\times n}$ and $G_1=\begin{bmatrix}
     G & 0
 \end{bmatrix}\in\mathbb \F_2^{k\times n}$.
We call $\mathcal{C}$ a \textit{UM simplex code}.
\end{definition}

For a UM simplex code equation~(\ref{eqG-totalUM}) becomes
\begin{equation}\label{eqG-totalUMsimplex}
	G_{\rm total} :=
	\begin{pmatrix}
      G & G & G & 0  &   &   &   &   &   &     \\
		&   & G & G & G & 0  &   & & &  \\
		&   &   &   & G & G & G & 0 & &  \\
		&   &   &   &   &   & \ddots & \ \ \ddots & \ \ \ddots  &  \\
		&   &   &   &   &   &   & G & G & G \ \ 0 \ \
	\end{pmatrix} ,
\end{equation}
with $G_{total} \in\F_2^{(s+1)k\times(s+2)n}$.

\begin{thm}
The distance of the block code generated by $G_{total}$ equals $3\cdot 2^{k-1}$. Moreover, for the column distances and the free distance of the corresponding UM simplex code as defined in Definition~\ref{simplexcc}, we have \[
d_0 = 2^{k}\;\;\AND \;\;d_{free}=d_j=3\cdot 2^{k-1} \; \FOR j \geq 1 .
\]
\end{thm}
\pf
The first column distance $d_0$ is the distance of the block code with generator matrix $[G \ G]$ which is obviously equal to $2^k$. Further,
$$
d_1 = \min \{ wt((c_0 , c_1)) |  (c_0 , c_1)= (u_0 , u_1)\begin{pmatrix}
		G & G  & G &   \\
		& & G & G &
	\end{pmatrix} \text{ with } u_0 \neq 0\}
    $$
       as $G$ is full row rank. Notice that the minimum weight can always be achieved considering $u_0 \neq 0$ and $u_1=0$ as selecting $u_1 \neq 0$ would never decrease the weight of the truncated codeword $(c_0, c_1)$. Then $d_1$ is the distance of the block code given by $[G \ G \ G]$ which is obviously $3 \cdot 2^{k-1}$. Analogous reasoning follows for all $d_j$, $j\geq 1$ and for $d_{free}$ which concludes the proof.
\pfend

\begin{rem}\label{remarkGc}
    The connection between the matrix $G_{total}$ as given by equation~(\ref{eqG-totalUMsimplex}) and the generator matrix $G_c$ of the block code ${\cal C}_2$ of the previous section is as follows: $G_c\otimes G$ equals the matrix $G_{total}$ with last $n/2$ zero columns deleted. Here $G_c\otimes G$ is the matrix that is obtained from $G_c$ by replacing all ones by $G$ and all zeros by the all-zero matrix of appropriate size.
\end{rem}

We now present the main theorem of this section which refers to the Easy Repair Property as defined in Definition~\ref{defEasy}.

\begin{thm}
 The UM simplex code as defined in Definition~\ref{simplexcc} has the Easy Repair Property.
\end{thm}

\pf

   Assume $c$ to be an erased node which can be repaired, i.e., $c=c_1+\cdots+c_x$ is the sum of the live nodes $c_1,\hdots,c_x$. Denote by $\mathfrak{g}$ and $\mathfrak{g}_i$ for $i=1,\hdots, x$, the columns of $G_{total}$ corresponding to the nodes. As we can leave away zero summands, we can assume that none of the live nodes corresponds to a zero column, i.e., to one of the last $2^k-1$ columns of $G_{total}$. As before, to prove the result it suffices to show that there is a pair $c_i,c_j$ with $i,j \in \{1,\hdots,x\}$ such that $\mathfrak{g}_i+\mathfrak{g}_j$ is a column of $G_{total}$.

    For $x=2$, we are done. Let $x\geq 3$ and $\mathfrak{g}=\mathfrak{g}_1+\cdots+\mathfrak{g}_x$. If there exist $i,j\in\{1,\hdots,x\}$, such that $\mathfrak{g}_i$ and $\mathfrak{g}_j$ belong to the same block of $2^k-1$ columns in $G_{total}$, then $\mathfrak{g}_i+\mathfrak{g}_j$ is a column of $G_{total}$.
       Hence, we can assume that all summands are from different blocks of size $2^k-1$.

    Considering the corresponding columns of $c=c_1+\cdots+c_g$ in $G_{total}$ we have
    $$
    \mathfrak{g}= \mathfrak{g}_1 + \mathfrak{g}_2 + \cdots + \mathfrak{g}_g.
    $$
    If $\mathfrak{g}_i$ is the $a_i$-th column of $G_{total}$ we assume that $a_{i+1}>a_i$, $i=1, ..., g-1$. Let
    $$
    \mathfrak{g}= \begin{pmatrix}
         0_{\ell k}\\ * \\ * \\ 0_{(s-\ell-1) k}
    \end{pmatrix}
    $$
    for some $\ell \in \{0,\hdots,s-1\}$.
    Let us consider two cases.

    Case 1: there exists a $\mathfrak{g}_i$ whose support contains indices  
  in a block of $k$ rows outside the support of $\mathfrak{g}$. Then, we have that $\mathfrak{g}_1$ and $\mathfrak{g}_2$ are of the form
    $$
    \begin{pmatrix}
         0_{\ell'  k}\\ g' \\  0_{(s-\ell') k}
    \end{pmatrix} \mbox{ and }
    \begin{pmatrix}
         0_{\ell' k}\\ g' \\ g' \\ 0_{(s-\ell'-1) k}
    \end{pmatrix}
    $$
    with $\ell' < \ell$ (if the smallest index in the support of $\mathfrak{g}_1$ is smaller than the smallest index in the support of $\mathfrak{g}$) or $\mathfrak{g}_{g-1}$ and $\mathfrak{g}_g$ are of the form
    $$
    \begin{pmatrix}
         0_{\ell' k}\\ g' \\ g' \\ 0_{(s-\ell'-1) k}
    \end{pmatrix} \mbox{ and }
    \begin{pmatrix}
         0_{(\ell' +1) k}\\ g' \\  0_{(s-\ell'-1) k}
    \end{pmatrix}
    $$
    with $\ell' > \ell$ (if the largest index in the support of $\mathfrak{g}_g$ is larger than the largest index in the support of $\mathfrak{g}$) where $g'$ is a column of $G$. In any case we have that $\mathfrak{g}_1 +\mathfrak{g}_2$ or $\mathfrak{g}_{g-1}+ \mathfrak{g}_g$ is a column of $G_{total}$.
    
    Case 2: If Case 1 does not hold then we must have $g=3$ and
    $$
   \underbrace{\begin{pmatrix}
         0_{\ell' k}\\ g \\ g \\ 0_{(s-\ell'-1) k}
    \end{pmatrix}}_{\mathfrak{g}} =\underbrace{ \begin{pmatrix}
         0_{\ell'  k}\\ g' \\  0_{(s-\ell') k}
    \end{pmatrix} }_{\mathfrak{g}_1} +
    \underbrace{\begin{pmatrix}
         0_{\ell' k}\\ g^{''} \\ g^{''} \\ 0_{(s-\ell'-1) k}
    \end{pmatrix}}_{\mathfrak{g}_2}
    +
     \underbrace{\begin{pmatrix}
         0_{(\ell' +1)  k}\\ g^{'''} \\  0_{(s-\ell'-1) k}
    \end{pmatrix}}_{\mathfrak{g}_3},
    $$
    for some columns $g'$, $g''$, $g'''$ of $G$, which implies that $g' = g^{'''}$ and therefore $\mathfrak{g}_1 + \mathfrak{g}_3$ is a column of $G_{total}$. This concludes the proof.
\pfend

We next investigate the parallel repair properties of UM simplex codes. To this end, we first need to present a technical theorem that requires some special notation.

Consider the matrix $G_{total}$ as given in equation~(\ref{eqG-totalUMsimplex}). Recall that equation~(\ref{c-total}) holds:
\begin{equation*}
    c_{\rm total} = u_{\rm total} G_{\rm total} .
\end{equation*}
We divide the set of nodes in each $c_i$ into two subsets as follows:
$$c_i = \begin{pmatrix}
    c_i^1 & c_i^2
\end{pmatrix}= \begin{pmatrix}
   c_{i,1}^1 & c_{i,2}^1 &\cdots & c_{i, \frac{n}{2}}^1 & c_{i,1}^2 & c_{i,2}^2 &\cdots & c_{i, \frac{n}
 {2}}^2
\end{pmatrix}= \begin{pmatrix}
    c_{i,1} & c_{i,2} &\cdots & c_{i,n}
\end{pmatrix},$$
where $c_i^1, c_i^2\in \F_2^{\frac{n}{2}}$ and $c_{i,j}^1, \ c_{i,j}^2,\ c_{i,j}\in \F_2$ denotes the $j$-th coordinate (node) of $c_i^1, \ c_i^2$ and $c_{i}$, respectively.

\begin{thm}\label{calc}
    Let $\mathcal{C}$ be a UM simplex $(n,k)$ code. Denote its set of nodes by $\{c_0,\dots,c_{s+\ell}  \}$, as before. Then,
    \begin{enumerate}
       \item each node $c_{0,j}$, $j\in \{ 1,\dots, n \}$, is equal to $1$ other node, i.e. allows for replication.
        \item each node $c_{0,j}$, $j\in \{ 1,\dots, n \}$, is related to
        %$\frac{n-1}{2} +1$
        $\frac{n}{2}=2^k-1$
        disjoint pairs of nodes.
         \item each node $c_{i,j}^1$ for $j\in\{1,\dots, \frac{n}{2}\}$, $i\in \{ 1,\dots, s+\ell \}$
         is related to $2^{k-1}$ disjoint pairs of nodes.
        \item each node $c_{i,j}^2$ for $j\in\{1,\dots, \frac{n}{2}\}$, $i\in \{ 1,\dots, s+\ell \}$, is related to
        $2^{k-1}+1$
        disjoint pairs of nodes.
        \item each node $c_{i,j}^1$ for $j\in\{1,\dots, \frac{n}{2}\}$, $i\in \{ 1,\dots, s+\ell \}$, is related to  %$(\frac{n-1}{2})^2 $ different sets of 3  nodes;
        $2^k-2$ disjoint sets of $3$
 nodes. Additionally, it is related to one disjoint set of $2$ nodes.      %\textcolor{cyan}{here the statement "is related to $\frac{n-1}{2} +1$ disjoint pairs of nodes" would be correct}
\item each node $c_{i,j}^2$ for $j\in\{1,\dots, \frac{n}{2}\}$, $i\in \{ 1,\dots, s+\ell \}$, is related to %$(\frac{n-1}{2})^2 $ different sets of 3  nodes;
        $2^k-2$ disjoint sets of $3$
 nodes. Additionally, it is related to two disjoint set of $2$ nodes.

        \item each node $c_{i,j}^1$ for $j\in\{1,\dots, \frac{n}{2}\}$, $i\in \{ 1,\dots, s+\ell \}$, is related to
$2^k$ disjoint sets of nodes where $2^{k-1}$ of these disjoint sets consist of $4$ nodes and the other $2^{k-1}$ sets consist of $2$ nodes.

\item each node $c_{i,j}^2$ for $j\in\{1,\dots, \frac{n}{2}\}$, $i\in \{ 1,\dots, s+\ell \}$, is related to
$2^k+2^{k-1}-1=d_{free}-1$ disjoint sets of nodes, where $2^{k}-2$ of these disjoint sets consist of $4$ nodes and the other $2^{k-1}+1$ sets consist of $2$ nodes.

       \item each node $c_{i,j}^1$ for $j\in\{1,\dots, \frac{n}{2}\}$, $i\in \{ 1,\dots, s+\ell \}$, is related to
$2^k+2^{k-1}-1=d_{free}-1$ disjoint sets of nodes, where $2^{k}-2$ of these disjoint sets consist of $5$ nodes, $2^{k-1}$ sets consist of $4$ nodes and one set consists of $2$ nodes.
                    \end{enumerate}
\end{thm}

\pf
\begin{enumerate}
  \item Follows directly from $c_{0,j}^1=c_{0,j}^2$ for $j=0,\hdots, \frac{n}{2}$.
    \item Without loss of generality assume that $j\leq\frac{n}{2}$, i.e. $c_{0,j}=c^1_{0,j}$. Since $G$ is the generator matrix of an $(\frac{n}{2},k)$ simplex block code, $c_{0,j}$ possesses $\frac{\frac{n}{2}-1}{2}$ disjoint repair pairs both in $c_0^1$ and $c_0^2$. Moreover, via $c^1_{0,j}=c^1_{1,j}+c^2_{1,j}$ one obtains an additional disjoint repair pair. In sum $c_{0,j}$ is related to $\frac{n}{2}$ disjoint pairs of nodes.
    \item Considering the submatrix $\begin{pmatrix}
        G & G & 0\\
        0 & G & G
    \end{pmatrix}$ of $G_{total}$, $c_{i,j}^1$ corresponds to a column of $\begin{pmatrix}
        G\\ G
    \end{pmatrix}$. Hence it possesses $2^{k-1}-1$ disjoint repair pairs inside $c_i^1$ plus the disjoint repair pair $c^1_{i,j}=c^2_{i-1,j}+c^2_{i,j}$.
    \item $c_{i,j}^2$ possesses $2^{k-1}-1$ disjoint repair pairs inside $c_i^2$ plus the two disjoint repair pairs $c^2_{i,j}=c^2_{i-1,j}+c^1_{i,j}$ and $c^2_{i,j}=c^1_{i+1,j}+c^2_{i+1,j}$.
    \item
       We fix an arbitrary node $c_{i,j}^1$. Then, for each $j_1\in\{1,\hdots,\frac{n}{2}\}\setminus\{j\}$, there exists $j_2\in\{1,\hdots,\frac{n}{2}\}\setminus\{j,j_1\}$, such that $c_{i,j}^1=c_{i,j_1}^1+c_{i-1,j_2}^2+c_{i,j_2}^2$ providing $2^{k}-2$ disjoint repair triples.
       Additionally, $c_{i,j}^1=c_{i-1,j}^2+c_{i,j}^2$ provides a disjoint repair pair.
    \item Follows directly from the previous part using that $c_{i,j}^2=c_{i-1,j}^2+c_{i,j}^1=c_{i+1,j}^1+c_{i+1,j}^2$, which gives us one additional disjoint repair pair.
    \item
        For $c_{i,j}^1$ there exist $2^{k-1}-1$ pairs $\{j_1,j_2\}$ such that $c_{i,j}^1=c_{i,j_1}^1+c_{i,j_2}^1=c_{i-1,j_1}^2+c_{i,j_1}^2+c_{i-1,j_2}^2+c_{i,j_2}^2$ giving as $2^{k-1}-1$ disjoint sets of nodes of size $4$. Additionally, we get another disjoint set of $4$ nodes via $c_{i,j}^1=c_{i-1,j}^2+c_{i,j}^2=c_{i-2,j}^2+c_{i-1,j}^1+c_{i+1,j}^1+c_{i+1,j}^2$.
    The $2^{k-1}$ disjoint sets of size $2$ are given by $c_{i,j}^1=c_{i,j_1}^1+c_{i,j_2}^1=c_{i-1,j}^2+c_{i,j}^2$.

    \item  For $c_{i,j}^2$ there exist $2^{k-1}-1$ pairs $\{j_1,j_2\}$ such that $c_{i,j}^2=c_{i,j_1}^2+c_{i,j_2}^2
    %=c_{i-1,j_1}^2+c_{i,j_1}^2+c_{i-1,j_2}^2+c_{i,j_2}^2
    $ giving us $2^{k-1}-1$ disjoint sets of nodes of size $2$. Additionally, we get another two disjoint sets of $2$ nodes via
$c^2_{i,j}=c^2_{i-1,j}+c^1_{i,j}$ and $c^2_{i,j}=c^1_{i+1,j}+c^2_{i+1,j}$.
    The $2\cdot(2^{k-1}-1)$ disjoint sets of size $4$ are given via
    $c_{i,j}^2=c_{i-1,j}^2+c_{i,j}^1=c_{i-1,j_1}^2+c_{i-1,j_2}^2+c_{i,j_1}^1+c_{i,j_2}^1$ and $c_{i,j}^2=c_{i+1,j}^1+c_{i+1,j}^2=c_{i+1,j_1}^1+c_{i+1,j_2}^1+c_{i+1,j_1}^2+c_{i+1,j_2}^2$.
    \item  We fix an arbitrary node $c_{i,j}^1$. Then, for each $j_1\in\{1,\hdots,\frac{n}{2}\}\setminus\{j\}$, there exists $j_2\in\{1,\hdots,\frac{n}{2}\}\setminus\{j,j_1\}$, such that $c_{i,j}^1=c_{i,j_1}^1+c_{i-2,j_2}^2+c_{i-1,j_2}^1+c_{i+1,j_2}^1+c_{i+1,j_2}^2$ providing $2^{k}-2$ disjoint repair sets of size $5$.
     $c_{i,j}^1=c_{i-1,j_1}^2+c_{i,j_1}^2+c_{i-1,j_2}^2+c_{i,j_2}^2$
    and
    $c_{i,j}^1=c_{i-2,j}^2+c_{i-1,j}^1+c_{i+1,j}^1+c_{i+1,j}^2$ provide $2^{k-1}$ disjoint sets of nodes of size $4$. Finally, $c_{i,j}^1=c_{i-1,j}^2+c_{i,j}^2$ gives a disjoint set of nodes of size $2$.
\end{enumerate}
This concludes the proof.\pfend

With the help of the above theorem, we study next the erasure patterns that the UM simplex code can cope with via parallel repair. More specifically, the next theorem provides results for parallel $r$-repairs up to $r=5$.

\begin{thm}
 Let $\mathcal{C}$ be a UM simplex $(n,k)$ code.
The following numbers of erasures can be recovered with parallel $r$-repairs:
\begin{enumerate}
    \item For $r=2$, any $2^{k-1}$ erasures can be recovered by parallel $r$-repairs.
    \item For $r=3$, any $2^{k}-1$ erasures can be recovered by parallel $r$-repairs.
    \item For $r=4$, any $2^{k}$ erasures can be recovered by parallel $r$-repairs.
    \item For $r=5$, any $2^k+2^{k-1}-1=d_{free}-1$ erasures can be recovered by parallel $r$-repairs.\end{enumerate}
\end{thm}

\pf
First note that $c_{0,j}$ has at least as many disjoint repair sets of given sizes as $c_{i,j}^2$ for $i\geq 1$.

An erasure pattern with a certain number of erasures $e$ can be corrected with parallel $r$-repairs
if all nodes are related to at least $e$ disjoint sets of nodes of size $\leq r$. Hence, all the results from this theorem follow immediately from Theorem~\ref{calc}.
\pfend

\begin{rem}
The final result of the previous theorem can be rephrased in the same vein as Theorem~\ref{c1-parallel} and Theorem~\ref{c2-parallel} as: for a UM simplex code any erasure pattern whose number of erasures is within the code's erasure-correcting capability allows for parallel $5$-repair.
\end{rem}

\section{Comparison}

In the previous sections we have introduced several codes and we have shown that each of them has the easy repair property. Given a fixed amount of data to be stored (which correspond to the dimension $k$) each of our constructions
will store it in a different number of nodes (the length $n$) and hence will yield a different rate $k/n$ and distance $d$. In this section we are setting out to compare some aspects of the different codes. To make a fair comparison,
we fix the dimension $k$ and then compute for each one of the constructions the corresponding rate $k/n$ and relative minimum distance $d/n$.
We note that it is desirable that codes of larger lengths $n$ have larger relative minimum distance $d/n$. This is analyzed next. 

\medskip

To compare the UM codes from the previous section to the block codes of section~\ref{sec_pun},
we fix the value of $k$, writing it as a multiple of an integer $x$. Then with $G_c\in\mathbb F_2^{x\times (2x+1)}$ and $G$ the generator matrix of a simplex code of dimension $k/x$ we consider the following matrix:

$$G_c\otimes G=\begin{pmatrix} G & G &  G &  & & \\
 &  & G & G & G &  & &\\
 &  &  &  &  G &  G & G & &   & \\
 &  &  &   &  & \ddots & \ddots & \ddots &\\
 &  &  &   &  &  & \ddots & \ddots & \ddots\\
 & &   &  &  &  &  &  G &  G & G
\end{pmatrix}\in\mathbb F_2^{k\times (2x+1)(2^{k/x}-1)} .$$

Thus $G_c\otimes G$ is the generator matrix of a block code of dimension $k$ and can be compared with other block codes of the same dimension. Note that the above matrix is the matrix of~(\ref{eqG-totalUMsimplex}) with the final block of zero columns removed.

The distance of the block code $UM_x^{(k)}$ generated by $G_c\otimes G$ is equal to $d_x'=3\cdot 2^{k/x-1}$ and it has length $n_x'=(2x+1)(2^{k/x}-1)$.
Note that $UM_k^{(k)}$ is exactly the code with generator matrix $G_c$ and hence, $\mathcal{C}_2$ does not need to be considered separately here.

Let $\mathcal{C}_0^{(k)}$ be a simplex code of dimension $k$ and let $S_k$ be its generator matrix. Similarly, let $\mathcal{C}_1^{(k)}$ be the code $\mathcal{C}_1$ from Section 4.1, where we additionally indicate its dimension $k$, and let $G_k$ be its generator matrix. For a fair comparison of these two codes with the codes $UM_x^{(k)}$ (all of the same dimension $k$), we also have to consider, for each divisor $x$ of $k$, the higher rate codes $\mathcal{C}_0^{(k,x)}$ with generator matrix $\diag(S_{k/x},\hdots,S_{k/x})\in\mathbb F_2^{k\times n_x}$ with $x$ matrices $S_{k/x}$ on the block diagonal, and the codes $\mathcal{C}_1^{(k,x)}$ with generator matrix $\diag(G_{k/x},\hdots,G_{k/x})\in\mathbb F_2^{k\times n_{x,p}}$ with $x$ copies of $G_{k/x}$. Note that $\mathcal{C}_i^{(k,1)}=\mathcal{C}_i^{(k)}$ for $i=0,1$.

It is immediate to see that all of these codes also have the easy repair property.
Moreover, $n_x=x(2^{k/x}-1)$, $d_x=2^{k/x-1}$, $n_{x,p}=x\left(\frac{k}{x}+\frac{k/x(k/x-1)}{s}\right)$, $d_{x,p}=\frac{k}{x}$ (the subscript $p$ stands for puncturing).

In the following we compare the performance of these classes of codes. Calculating the ratios $\frac{d}{n}$, one obtains:
\begin{align}\label{ratios}
    \frac{d_x}{n_x}\approx \frac{1}{2x},\quad \frac{d'_x}{n'_x}\approx \frac{3}{2(2x+1)},\quad \frac{d_{x,p}}{n_{x,p}}\approx \frac{1}{x+(k-x)/2}
\end{align}

Therefore, comparing classical simplex block codes and UM simplex convolutional codes, one gets

\begin{align}
    \frac{n'_x}{n_x}\approx 2; \quad, \frac{d'_x}{d_x}\approx 3; \quad n'_x>n_x
\end{align}

This means that we can cover with the UM simplex codes rates that classical simplex block codes cannot cover and we observe that, considering all constructions together, $\frac{d}{n}$ is increasing if $\frac{k}{n}$
is decreasing. We will illustrate this with the following tables of example values, where we also observe that it highly depends on $k$ how the codes $\mathcal{C}_1^{(k,x)}$ compare to the other codes (\eqref{ratios} shows that they perform better for small $k$).

\begin{center}
$k=4$:\\
\vspace{2mm}
\begin{tabular}{|c|c| c| c|}
 \hline
 Code & $n$ & $d$ & $d/n$ \\ [0.5ex]
 \hline\hline
 $\mathcal{C}^{(4)}_0$ & 15 & 8 & $\approx 1/2$ \\
 \hline
 $UM_2^{(4)}$ & 15 & 6 & $\approx 1/3.5$ \\
 \hline
 $\mathcal{C}^{(4)}_1$ & 10 & 4 & $\approx 1/2.5$ \\
 \hline
 $UM_2'^{(4)}$ & 9 & 4 & $\approx 1/2$ \\
 \hline
 $UM_4^{(4)}$ & 9 & 3 & $\approx 1/3$ \\
 \hline
 $\mathcal{C}_0^{(4,2)}=\mathcal{C}_1^{(4,2)}$
  & 6 & 2 & $\approx 1/3$ \\ [1ex]
 \hline
\end{tabular}
%\caption{Comparison of codes with $k=4$ }
\end{center}

From the table we see that the code $UM_2^{(4)}$, which has generator matrix
$\begin{pmatrix}
    G & G & G & 0 & 0\\
    0 & 0 & G & G & G
\end{pmatrix}$ with $G=\begin{pmatrix}
    1 & 1 & 0\\
    0 & 1 & 0
\end{pmatrix}$,
is the only code in the list for which there exists a code with the same rate
but better distance, namely $\mathcal{C}^{(4)}_0$. The reason for this is that the UM codes do not perform well for very small $k$ as the repeated columns in the blocks $\begin{pmatrix}
    G\\ 0
\end{pmatrix}$ and $\begin{pmatrix}
    0\\ G
\end{pmatrix}$ have a relatively large effect on the rate.
For comparison, we added to the above ($k=4$) table the code $UM_2'^{(4)}$ with generator matrix defined as
$\begin{pmatrix}
     G & G & 0 \\
     0 & G & G
\end{pmatrix}$.
It is clear from the table that this code is preferable over $UM_2^{(4)}$.

\newpage

\begin{center}
$k=6$:\qquad \qquad\qquad\qquad\qquad\qquad\qquad$k=8:$\\
\vspace{2mm}
\begin{tabular}{|c|c| c| c|}
 \hline
 Code & $n$ & $d$ & $d/n$ \\ [0.5ex]
 \hline\hline
 $\mathcal{C}_0^{(6)}$ & 63 & 32 & $\approx 1/2$ \\
 \hline
 $UM_2^{(6)}$ & 35 & 12 & $\approx 1/3$ \\
 \hline
 $\mathcal{C}_1^{(6)}$ & 21 & 6 & $ 1/3.5$ \\
 \hline
 $UM_3^{(6)}$ & 21 & 6 & $ 1/3.5$ \\
 \hline
 $\mathcal{C}_0^{(6,2)}$ & 14 & 4 & $ 1/3.5$ \\
 \hline
  $UM_6^{(6)}$ & 13 & 3 & $\approx 1/4$ \\
 \hline
 $\mathcal{C}_1^{(6,2)}$ & 12 & 3 & $ 1/4$\\
 \hline
 $\mathcal{C}_0^{(6,3)}=\mathcal{C}_1^{(6,3)}$ & 9 & 2 & $ 1/4.5$ \\ [1ex]
 \hline
\end{tabular}\quad
\begin{tabular}{|c|c| c| c|}
 \hline
 Code & $n$ & $d$ & $d/n$ \\ [0.5ex]
 \hline\hline
 $\mathcal{C}_0^{(8)}$ & 255 & 128 & $\approx 1/2$ \\
 \hline
 $UM_2^{(8)}$ & 75 & 24 & $\approx 1/3$ \\
 \hline
 $\mathcal{C}_1^{(8)}$ & 36 & 8 & $1/4.5$ \\
 \hline
 $\mathcal{C}_0^{(8,2)}$ & 30 & 8 & $\approx 1/4$ \\
 \hline
 $UM_4^{(8)}$ & 27 & 6 & $ 1/4.5$ \\
 \hline
 $\mathcal{C}_1^{(8,2)}$ & 20 & 4 & $1/5$ \\
  \hline
 $UM_6^{(8)}$ & 17 & 3 & $\approx 1/6$
  \\
  \hline
 $\mathcal{C}_0^{(8,4)}=\mathcal{C}_1^{(8,4)}$ & 12 & 2 & $1/6$
  \\ [1ex]
 \hline
\end{tabular}\quad
\end{center}

We observe that with only very few exceptions $d/n$ decreases when the rate increases, showing that all presented constructions with the easy repair property have essentially the same relative erasure-correcting capability. Hence, their main difference is the code rate and depending on the situation one can choose which rate is preferred.

In all our tables, only $UM_2^{(4)}$, which can be replaced by the well-performing code $UM_2'^{(4)}$ as explained above, and $\mathcal{C}_1^{(k)}$ for $k=8$ do not perform well in comparison to the other codes. This also fits with the observation we made above, i.e. that \eqref{ratios} indicates that $\mathcal{C}_1^{(k)}$ will only perform well for small values of $k$.

\section{Conclusions}\label{sec_conclusions}

In this paper we presented several families of codes over fields of characteristic $2$ with the property that all correctable erasure patterns can be recovered with easy repairs. Hence, these families of codes are highly suitable for efficient erasure coding in multiple-erasure settings within Networked Distributed Storage Systems. They allow for easy encoding and decoding, requiring only addition operations in $\mathbb F_q$.

All code constructions are related to the simplex generator matrix with an eye on feeding off the natural easy repair properties of the simplex code.  We treat two different scenarios: when one encodes a fixed length data vector and when one encodes an arbitrary long sequence of vectors of data. For the first case, we study the properties of simplex codes as well as certain puncturings of these codes. For the second scenario, we propose a novel construction of unit memory convolutional codes built upon two blocks of generator matrices of simplex codes. These codes inherit some of the properties of simplex block codes and allow for sliding window easy repairs.

We investigate the availability properties of all presented constructions, deriving the number of erasures that can be recovered in parallel using $r$-repairs. It turns out that the convolutional code case is more involved than the block code scenario and different values for $r$ up to $r=5$ have to be considered to reach the erasure correcting capability of the code.
Finally, we compare all the presented constructions with respect to code rate and minimum distance. This comparison shows no significant differences in the performance, meaning that overall we are able to provide well-performing codes with the easy repair property for a large variety of different code rates.

\section{Acknowledgments}
The first author would like to thank the University of Z\"urich as well as the Technical University of Ilmenau for the hospitality that enabled research visits.
The second author would like to thank the University of Alicante for their invitation and hospitality.
The second author is supported by the German research foundation, project number 513811367.
 The third author is partially supported by the Spanish I+D+i project PID2022-142159OBI00 of the Ministerio de Ciencia e Innovación and I+D+i project CIAICO/2022/167 of the
Generalitat Valenciana.

\end{document}